\RequirePackage{fix-cm}
\documentclass[twocolumn]{svjour3}          
\smartqed  
\usepackage{graphicx}
 \usepackage{color} 
 \usepackage{amsmath}
\begin{document}

\title{Detectability of Rocky-Vapour Atmospheres on Super-Earths with Ariel}
\subtitle{}


\author{Yuichi Ito         \and
        Quentin Changeat    \and 
        Billy Edwards
        \and 
        Ahmed Al-Refaie   
        \and 
        Giovanna Tinetti
        \and 
        Masahiro Ikoma
}


\institute{Yuichi Ito \at
              Department of Physics and Astronomy, University College London, London, UK
              \email{yuichi.ito@ucl.ac.uk}           
              }
\date{Accepted: 22. Jan. 2021}

\maketitle

\begin{abstract}

Ariel will mark the dawn of a new era as the first large-scale survey characterising exoplanetary atmospheres with science objectives to address fundamental questions about planetary composition, evolution and formation. In this study, we explore the detectability of atmospheres vaporised from magma oceans on dry, rocky Super-Earths orbiting very close to their host stars. The detection of such atmospheres would provide a definitive piece of evidence for rocky planets but are challenging measurements with currently available instruments due to their small spectral signatures. However, some of the hottest planets are believed to have atmospheres composed of vaporised rock, such as Na and SiO, with spectral signatures bright enough to be detected through eclipse observations with planned space-based telescopes. 
In this study, we find that rocky super-Earths with a irradiation temperature of 3000~K and a distance from Earth of up to 20~pc, as well as planets hotter than 3500~K and closer than 50~pc, have SiO features which are potentially detectable in eclipse spectra observed with Ariel.


\end{abstract}

\section{Introduction}
\label{intro}

Currently over 1000 exoplanets have been detected whose radii are less than twice that of the Earth. About 50~\% of those planets have irradiation temperatures at substellar-points high enough to melt and vaporise rock. Based on some planetary formation and evolution models \cite{Owen+2017,Jin+2018}, which have reproduced the distribution of discovered exoplanets \cite{Fluton+2017}, most of the close-in small exoplanets are likely bare rocky planets as they lost their primordial hydrogen-rich atmospheres due to photo-evaporation.
Close-in terrestrial planets, such as CoRoT-7~b, are likely to have secondary atmospheres vaporized from their magma oceans \cite{Valencia+2010}.
Gas-melt equilibrium calculations \cite{Schaefer+2009,Miguel+2011,Ito+2015} have shown that the main constituents of such atmospheres are expected to be gas-phase Na, K, Fe, Si, O, O$_2$ and SiO on magma oceans without highly volatile elements such as H, C, N, S, and Cl (i.e., volatile-free magma oceans). We will refer to a close-in rocky planet as a hot rocky exoplanet (HRE, hereafter) and such a rocky-vapour atmosphere as a mineral atmosphere.
Thus, identifying the atmospheric constituents could give constraints on their magma compositions which are key to understanding the bulk compositions and also formation processes of the planets \cite{Morbidelli+2016}.

In particular, the detection of a mineral atmosphere would provide a definitive piece of evidence for rocky planets. A previous study estimated the emission spectra of such mineral atmospheres on HREs \cite{Ito+2015} and showed that those spectra exhibit prominent features of Na, K, Fe and SiO, the strength of which increases with the planetary irradiation temperature. Among them, the SiO features at 4 and 10~$\mu$m were found to be most prominent.

In the coming decade, we  expect to move into a new era of characterisation for exoplanets. The next-generation of instruments will access the atmospheres of planets which were far out of reach with current telescopes, not only observing the atmospheric properties of large, gaseous planets but also unveiling the composition of terrestrial rocky exoplanets. Dedicated missions, such as  Ariel \cite{Tinetti+2018}, and large space observatories, such as the James Webb Space Telescope (JWST) \cite{Gardner+2006}, will provide increased sensitivity and spectral coverage. Ariel is expected to characterise around 1000 exoplanets as a dedicated survey mission for the atmospheric spectroscopy with a planned launch date of 2029 \cite{Tinetti+2018}. 

The Ariel mission will study a diverse population of exoplanets \cite{Edwards+2019} and its simultaneous wavelength coverage, from 0.5-7.8 $\mu m$, provides the ability to detect a wide variety of molecular features. In particular, the SiO feature around 4~$\mu$m is within the spectral range of Ariel and therefore could potentially be an excellent tracer to  identify the presence of a mineral atmosphere. We note that SiO gas is expected to be present in the hydrogen-rich atmospheres of hot Neptunes as well \cite{Lothringer+2018}. The ambiguity can however be informed by knowledge of the planetary mass, which can be obtained independently from radial velocity measurements.

This study aims to assess the detectability of the 4-$\mu$m SiO feature in eclipse observations obtained with Ariel. As a case study, we focus our attention upon 55~Cnc~e as it is one of the best candidates from the current list of Ariel targets \cite{Edwards+2019}. We note that, as some observations suggest atmospheric compositions of 55~Cnc~e that differ from a mineral atmosphere \cite{Gillon+2012,Bourrier+2018}, 55~Cnc~e's parameters are chosen purely as an example.
Additionally, we investigate which kind of planets potentially have detectable SiO features in eclipse observations with Ariel, focusing on two parameters: irradiation temperature and distance from Earth. These two parameters are essential for the detection of HREs, as SiO emission features increase with the former, while the signal decreases with the latter  \cite{Ito+2015}.

The remainder of this paper is organised as follows. In Section~\ref{sec:model}, we describe our atmospheric model and numerical setup. In Section~\ref{sec:det}, we show the detectability of mineral atmospheres in eclipse observations with Ariel. Next, we discuss the number of planets potentially having a detectable signature of mineral atmospheres and future perspective for the observation with Ariel in Section~\ref{sec:dis}. Finally, we summarise our results in Section~\ref{sec:con}.

\section{Methods} \label{sec:model}
To investigate the potential eclipse spectra and simulated Ariel observation of a mineral atmosphere for a 55 Cnc e-like planet, we assume $R_p$ = 1.99~R$_\oplus$, $M_p$ = 8.09~M$_\oplus$, $a$ = 0.0155 AU, $R_*$ =0.94~R$_\odot$, $T_*$ = 5200~K as the planetary radius, planetary mass, semi-major axis, stellar radius and the temperature of the host star, respectively \cite{Dragomir+2014,Ehrenreich+2012,Nelson+2014}. 
In addition, we consider mock-planets for wide ranges of irradiation temperatures at substellar-points. Then, we assume 
$R_p$ = 2~R$_\oplus$, $M_p$ = 10~M$_\oplus$, $R_*$ = 1~R$_\odot$ with the stellar emission modelled as a blackbody with a temperature of 6000~K.
The irradiation temperature at a substellar-point is given by 
\begin{equation}
    T_{\mathrm{irr}}^4=\frac{R_*^2}{a^2}T_*^4.
    \label{eq:Tirr}
\end{equation}
Using this equation, we calculate the resulting spectra from planets located at different semi-major axis in our simulations.

\begin{figure}
\includegraphics[width=0.45\textwidth]{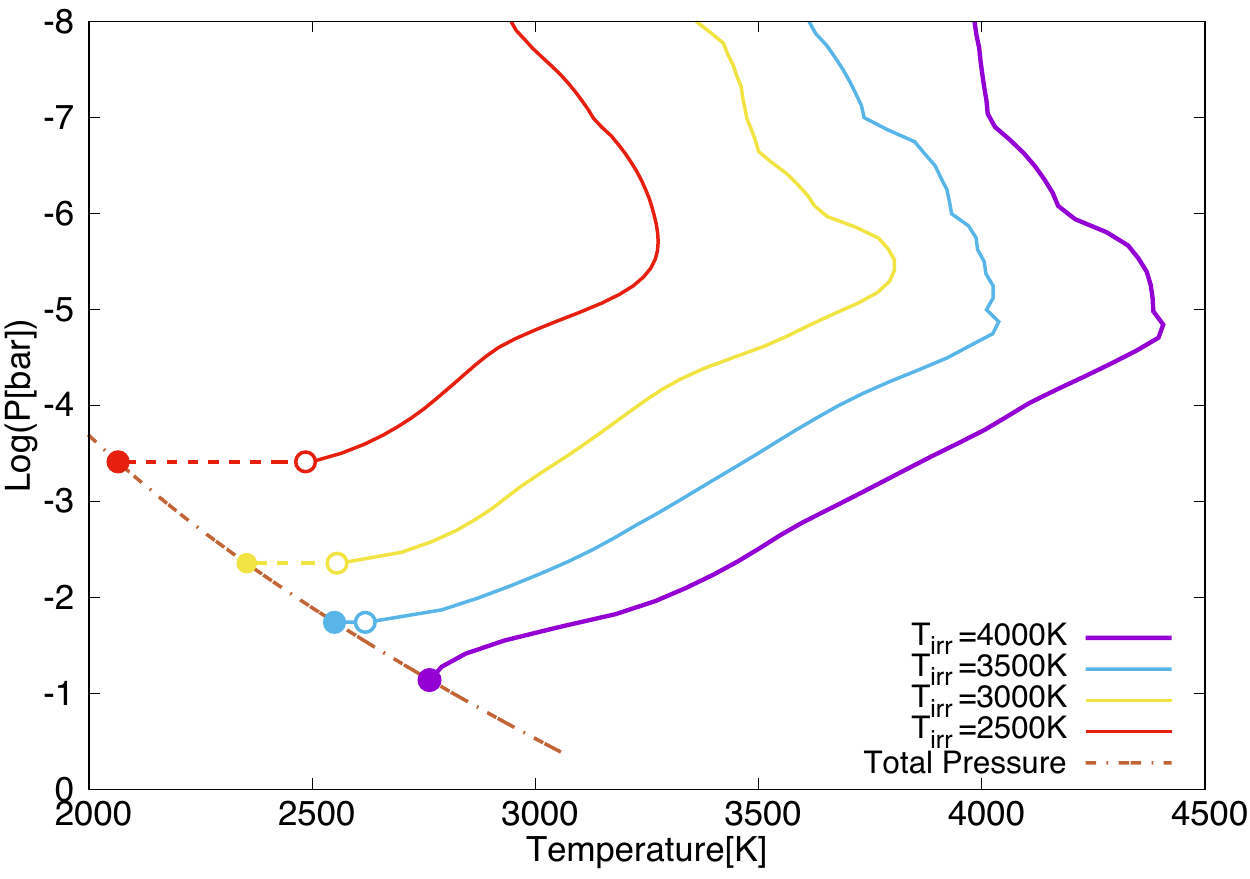}
\caption{
Temperature-pressure profile of a mineral atmosphere on top of the BSE magma of a hot rocky exoplanet with a radius of 2~R$_\oplus$ and a mass of 10~M$_\oplus$ for four choices of the substellar-point irradiation temperature, $T_\mathrm{irr}=$ 2500~K, 3000~K, 3500~K and 4000~K. The solid lines show the dayside-averaged profiles (i.e., the cosine of the stellar-light zenith angle is 0.5, see Eq.(12) of \cite{Ito+2015}).  The open circles show the temperatures at the bottom of the atmosphere and the filled circles show the temperatures at the planet's surface. The orange dotted line represents the total vapor pressure for the BSE composition, which corresponds to the planetary surface.
    }
\label{fig:HSE_TP}
\end{figure}

\begin{figure}
 \begin{minipage}{0.45\textwidth}
    ($a$) $T_\mathrm{irr}=$ 2500~K
    \begin{center}
  \includegraphics[width=\textwidth]{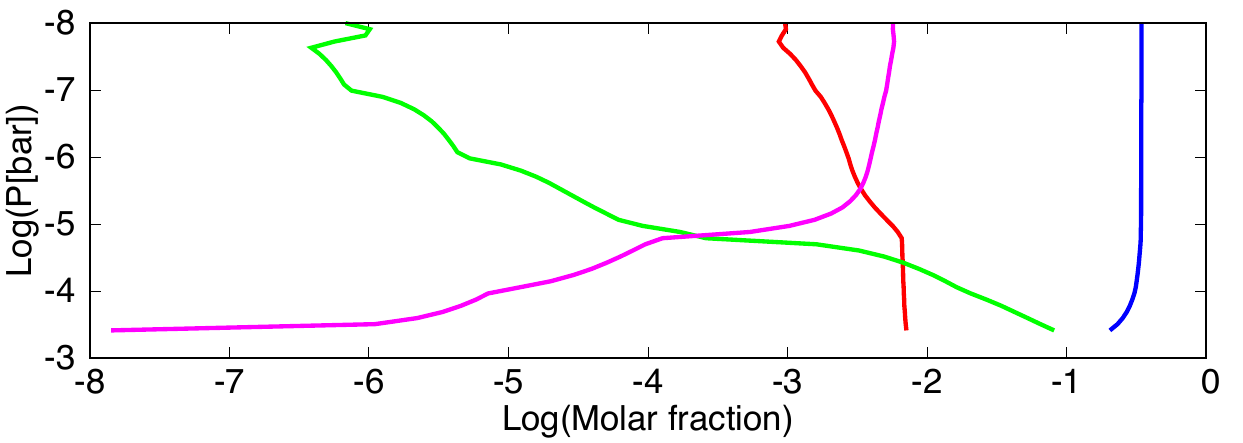}
   \end{center}
 \end{minipage}
 \begin{minipage}{0.45\textwidth}
    ($b$) $T_\mathrm{irr}=$ 3000~K
    \begin{center}
  \includegraphics[width=\textwidth]{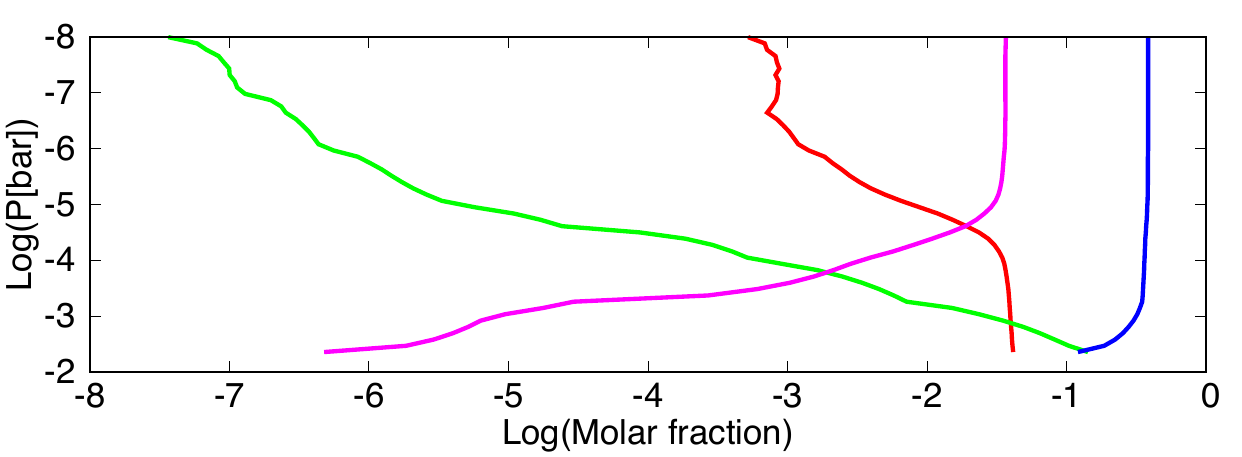}
   \end{center}
 \end{minipage}
  \begin{minipage}{0.45\textwidth}
    ($c$) $T_\mathrm{irr}=$ 3500~K
    \begin{center}
  \includegraphics[width=\textwidth]{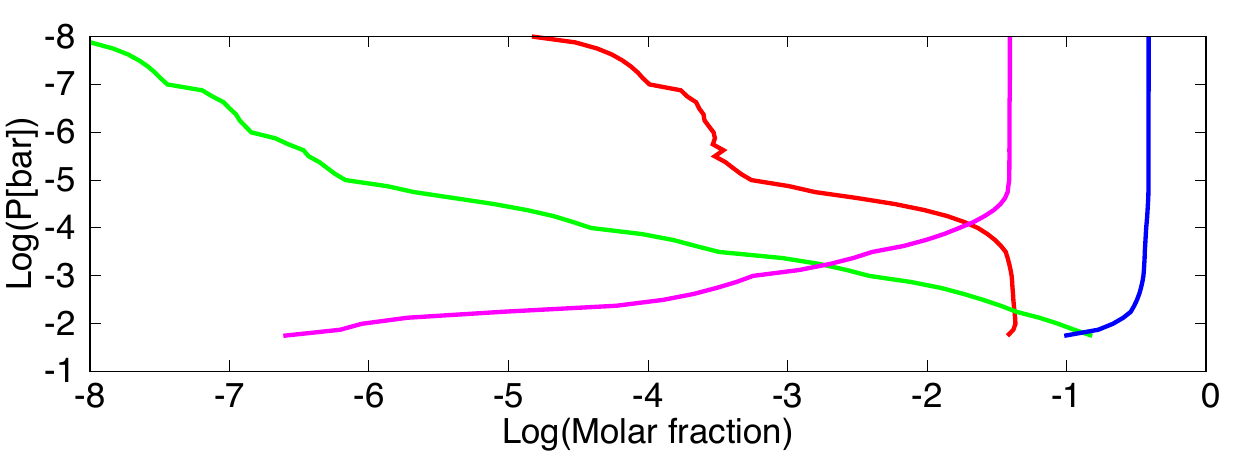}
   \end{center}
 \end{minipage}
  \begin{minipage}{0.45\textwidth}
    ($d$) $T_\mathrm{irr}=$ 4000~K
    \begin{center}
  \includegraphics[width=\textwidth]{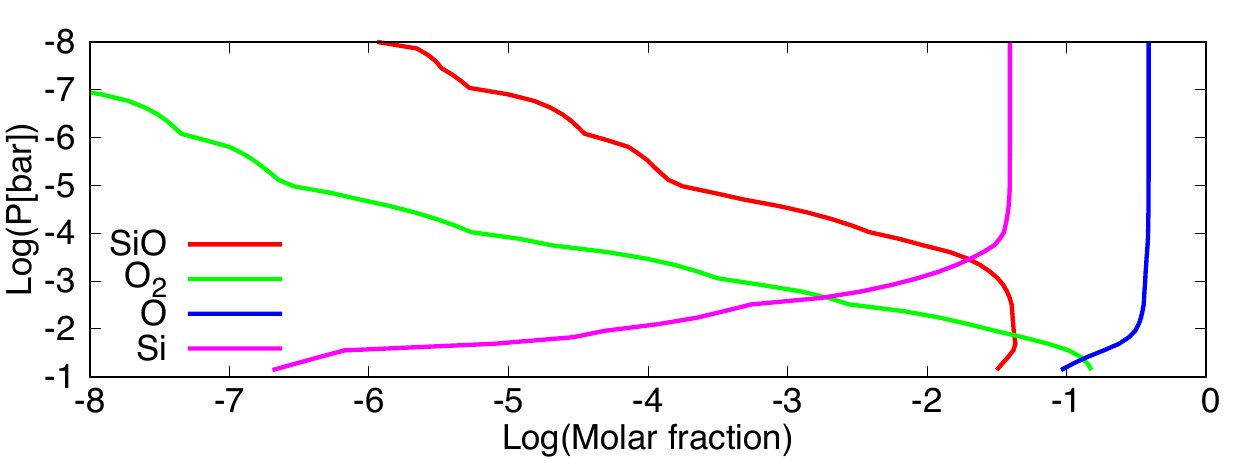}
   \end{center}
 \end{minipage}
\caption{Vertical distributions of SiO and its chemically-related gases in the mineral atmosphere. Molar fractions of SiO (red), O$_2$ (green), O (blue) and Si (magenta) are shown as functions of pressure for four choices of the substellar-point irradiation temperature, (a)~$T_\mathrm{irr}=$ 2500~K, (b)~3000~K, (c)~3500~K and (d)~4000~K.
    }
\label{fig:atom_comp}
\end{figure}

We use the model for mineral atmospheres from \cite{Ito+2015}  to estimate the  eclipse depth spectra at high resolution. 
The model calculates the gas-melt equilibrium composition of the atmosphere with the vapor pressure of magma and the vertical structure in hydrostatic/~radiative/~chemical equilibrium.
In the model, given a magma composition, an atmospheric composition and a pressure in chemical equilibrium are estimated using two open codes, MELTS \cite{Ghiorso+95,Asimow+98} for the melt-phase chemistry and NASA CEA code \cite{Gordon+96} for the gas-phase chemistry, while a radiative equilibrium solution is obtained based on a two-stream radiative transfer calculation \cite{Toon+89}. For the radiative property of the atmosphere, the absorption lines with Voigt-profiles of major gas species including SiO, Na, K, Fe, O$_2$, O and Si in the atmosphere are considered using the numerical tables of their opacities with several grids of temperatures and pressures calculated in \cite{Ito+2015} and we refer the reader to this study for more information.
In agreement with the previous study \cite{Ito+2015}, we assume a volatile-free Bulk Silicate Earth (BSE) composition as the magma composition and zero as the albedo value. As the albedo of  molten silicates is as low as $\leq$0.1 \cite{Essack+2020,Modirrousta-Galian+2020a},  we neglect this effect as it would not significantly affect the surface temperature. If one considers optically thin atmospheres and the albedo value, $A$, of 0.1, the temperature change is only a few percent as $T \propto [1-A]^{1/4}\sim$0.97 at A=0.1.

As examples of the structures of mineral atmospheres, the resulting temperature-profiles and the distributions of SiO and gases  chemically related with SiO for the 2~R$_\oplus$ mock-planets with $T_\mathrm{irr}=$ 2500~K, 3000~K, 3500~K and 4000~K are shown in Fig.~\ref{fig:HSE_TP} and Fig.~\ref{fig:atom_comp}, respectively. One can see the thermal inversion structures of the mineral atmospheres. Also, the decrease in the SiO abundance is occurring in low pressure regions below $\sim10^{-3}$~bar. This is because the diatomic molecules such as SiO and O$_2$ are thermally dissociated due to the high temperature in the low pressure region while the molar fraction of the other major species such as Na, K and Fe hardly change with pressure (see also Fig.5 of \cite{Ito+2015}). Also, the thermal inversion is a consequence of heating by the strong UV-ray absorption of SiO and the reduction of the heating due to the thermal dissociation of SiO in the low pressure region. Thus, the temperature peak increases with $T_{\mathrm{irr}}$ and there is more SiO dissociation at higher $T_{\mathrm{irr}}$. These behaviors with irradiation temperature have been already explored in details in the previous study \cite{Ito+2015}.

To simulate Ariel performances, we use the radiometric model, ArielRad, described in \cite{Mugnai+2019}. ArielRad provides the  signal to noise as a function of wavelengths at the spectral resolving power of Ariel. The high spectral  resolution spectrum can be convolved to the Ariel spectral resolving power using TauREx3  \cite{Al-refaie+2019}. 

In a mineral atmosphere, SiO presents large features around 4 $\mu$m and 10 $\mu$m \cite{Ito+2015} . Unfortunately, Ariel only covers the wavelength range from 0.5 $\mu$m to 7.8 $\mu$m, which means that the large 10 $\mu$m SiO feature lies outside its detection range. We can only rely on the 4.5 $\mu$m feature to assess the detection strength of SiO. We note that JWST MIRI LRS \cite{Kendrew+2015} will cover this 10 $\mu m$ feature and future works will study its capability for detecting mineral atmospheres. For our assessment of the SiO detection, we calculate the strength of the SiO signal ($S_{SiO}$) in the 4-5.3 $\mu$m window using the following formula:
\begin{equation}
    S_{\rm SiO} = \underset{\lambda}{\mathrm{max}} \left(\frac{S(\lambda) - BB_{\mathrm{ref}}(\lambda)}{\sigma}\right),
\label{eq:S_SiO}
\end{equation}
where $\lambda$ is a wavelength in the range 4 $\mu$m- 5.3 $\mu$m, $S$ is the observed emission spectrum at Ariel resolution, $BB_{\mathrm{ref}}$ is the reference blackbody calculated from the brightness temperature at the maximum of S for $\lambda$ = 3.7 $\mu$m and  $\lambda$ = 5.6 $\mu$m and $\sigma$ is the average Ariel noise in the window 4-5.3 $\mu$m.
In essence, equation \ref{eq:S_SiO} gives us a metric to compare the SiO signal at 4.5 $\mu$m with the signal of a similar planet without SiO, expressed in terms of the corresponding simulated noise of an Ariel observation.

\section{Detectability with Ariel} \label{sec:det}
In this section, we present the results of our study. In the first place, we show an example for the exoplanet with 55 Cnc e's property. Then, we extend our study to the entire parameter space and constrain the regions wherein terms of planet temperature and star distance from Earth; it will be possible to detect SiO.

\subsection{Case of 55 Cnc e like planet} \label{ssec:55}

\begin{figure}
\includegraphics[width=0.45\textwidth]{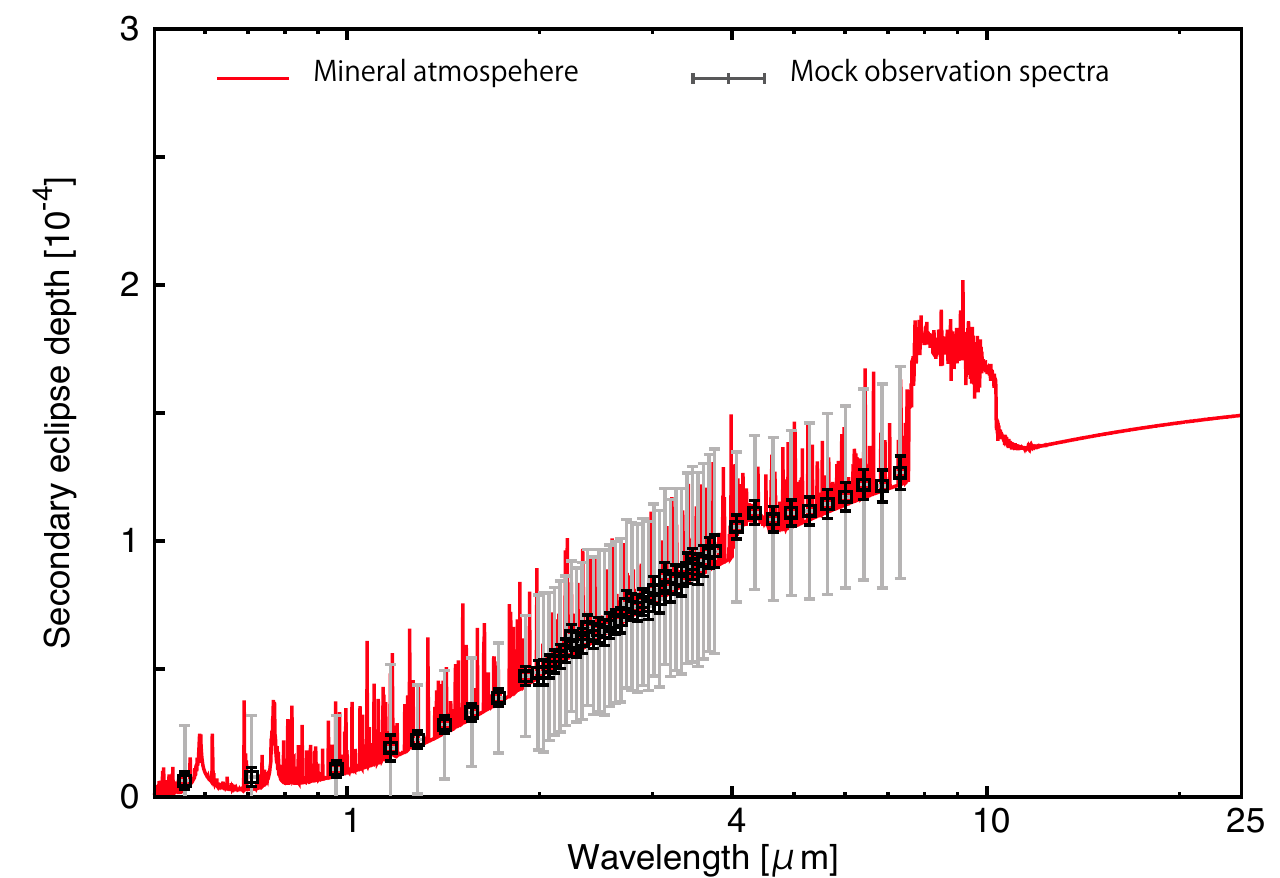}
\caption{Theoretically predicted (red line) and mock (black and grey bars) spectra of secondary eclipse depth for a 55-Cnc-e-like transiting planet with a mineral atmosphere.
The grey and black bars in the mock spectra represent cumulative errors estimated for 1 and 40 eclipse observations with Ariel, respectively.
    }
\label{fig:55cnc_spectra}
\end{figure}

We first take the case of 55 Cnc e like planet.
The resulting spectra are shown in Figure~\ref{fig:55cnc_spectra}. The theoretical spectra include line features induced by Na ($\sim$0.6~$\mu$m), K ($\sim$0.8~$\mu$m), SiO ($\sim$4~$\mu$m and $\sim$10~$\mu$m) and Fe (mainly, $\leq$0.6~$\mu$m; see \cite{Ito+2015} for further details) with the continuum emission feature from the surface. The value of the secondary eclipse depth around 4~$\mu$m is  $\sim$100~ppm while the difference between the SiO feature and the reference Black Body spectrum at the maximum of $S$ for $\lambda$ = 3.7 $\mu$m and  $\lambda$ = 5.6~$\mu$m is only $\sim$7~ppm. On the other hand, the Ariel observational noises at 4.3~$\mu$m are about 30~ppm and 5~ppm with 1 and 40 secondary eclipses, respectively.

\begin{figure}
\includegraphics[width=0.45\textwidth]{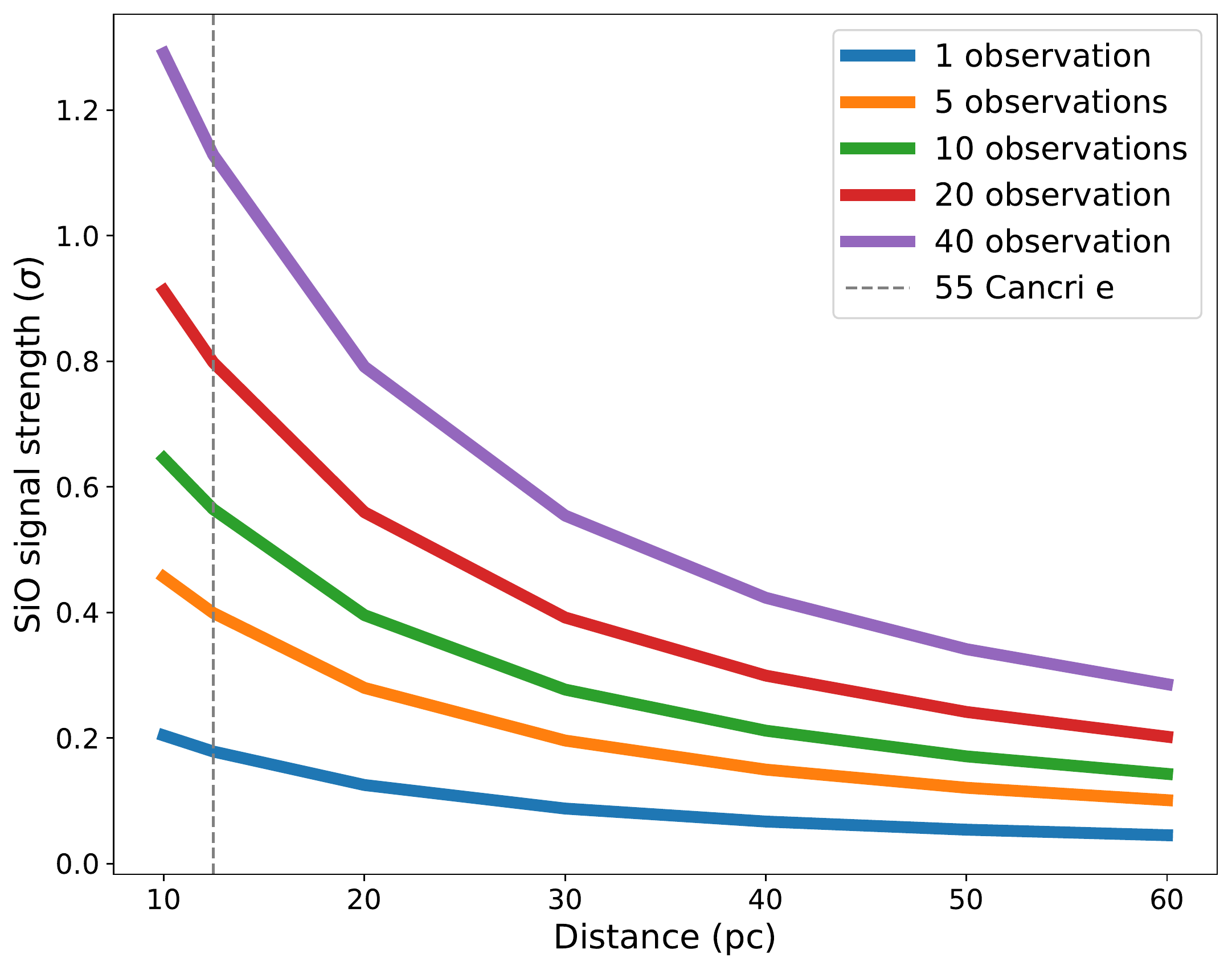}
\caption{Simulated SiO signal strength ($S_{\rm SiO}$) for a planet like 55 Cnc e located at different distances for Ariel observations.}
\label{fig:55cnc_distance}
\end{figure}

Figure~\ref{fig:55cnc_distance} shows the SiO detectability in a 55 Cnc e like planet as a function of star distance (pc) from Earth for 1, 5, 10, 20 and 40 eclipses observations with Ariel.
As one can see, even with 20 eclipse observations, the SiO signal is very difficult: 
$S_{\rm SiO}$ is less than or equal to 1 at all distances. However, although it may be unrealistic to stack so many observations, the accumulation of 40 secondary eclipses with Ariel could lead to the detection of the 4-$\mu$m SiO feature, provided the planet has a mineral atmosphere. This difficulty is because the 4-$\mu$m SiO feature is small for the substellar-point equilibrium temperature of 2700~K, which corresponds to that of 55 Cnc e. The detectability of this signature for hotter planets is examined in Section~\ref{ssec:ps}.
Also, for a 55 Cnc e like planet, even if the detection of SiO is difficult, Ariel observations with the reasonable number of eclipses may be quite helpful to distinguish the mineral atmosphere and volatile-rich atmospheres. The possibility is discussed in Section~\ref{ssec:dis0}.

\subsection{Parameter space of detection}\label{ssec:ps}
\begin{figure}
  \begin{minipage}{0.45\textwidth}
    ($a$) High resolution spectra 
    \begin{center}
  \includegraphics[width=\textwidth]{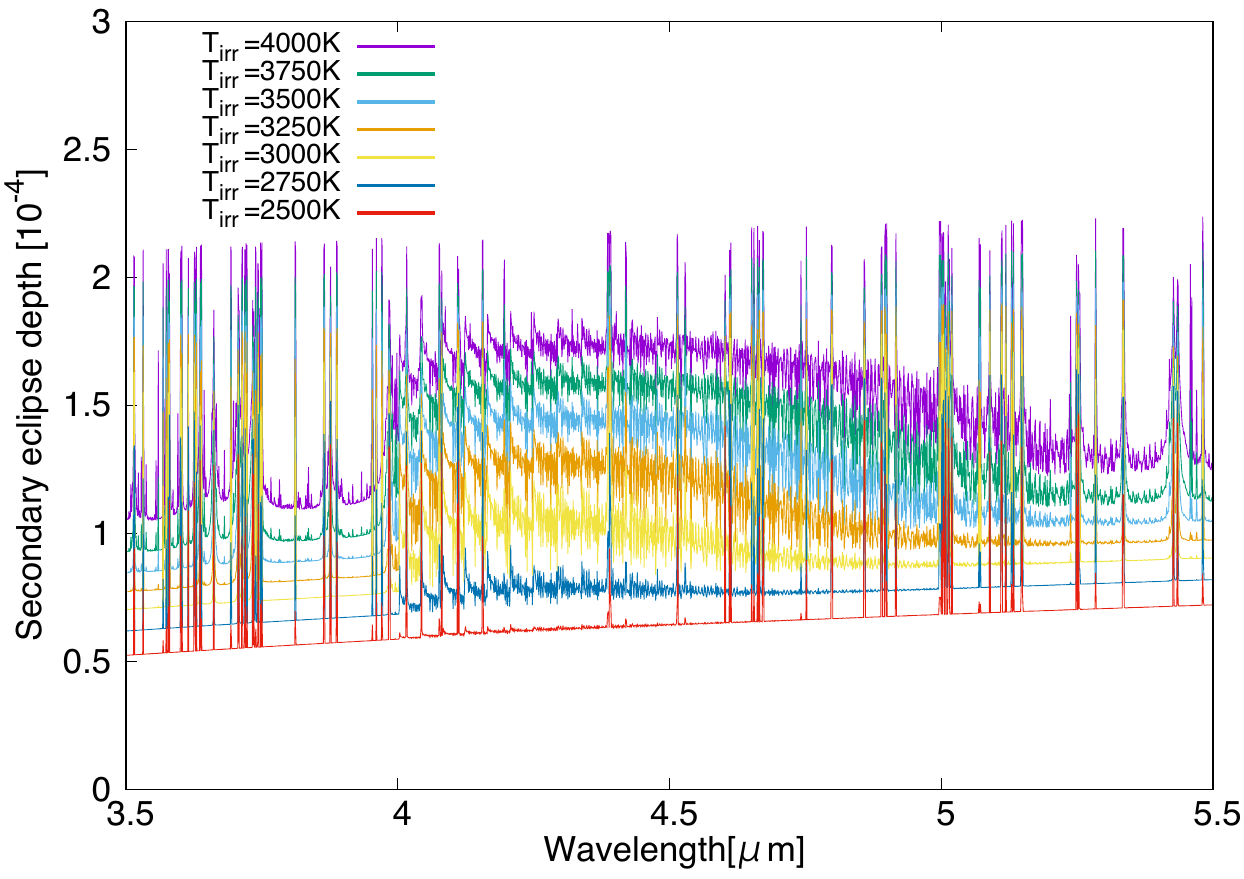}
    \end{center}
 \end{minipage}
  \begin{minipage}{0.45\textwidth}
    ($b$) Ariel resolution spectra 
    \begin{center}
  \includegraphics[width=\textwidth]{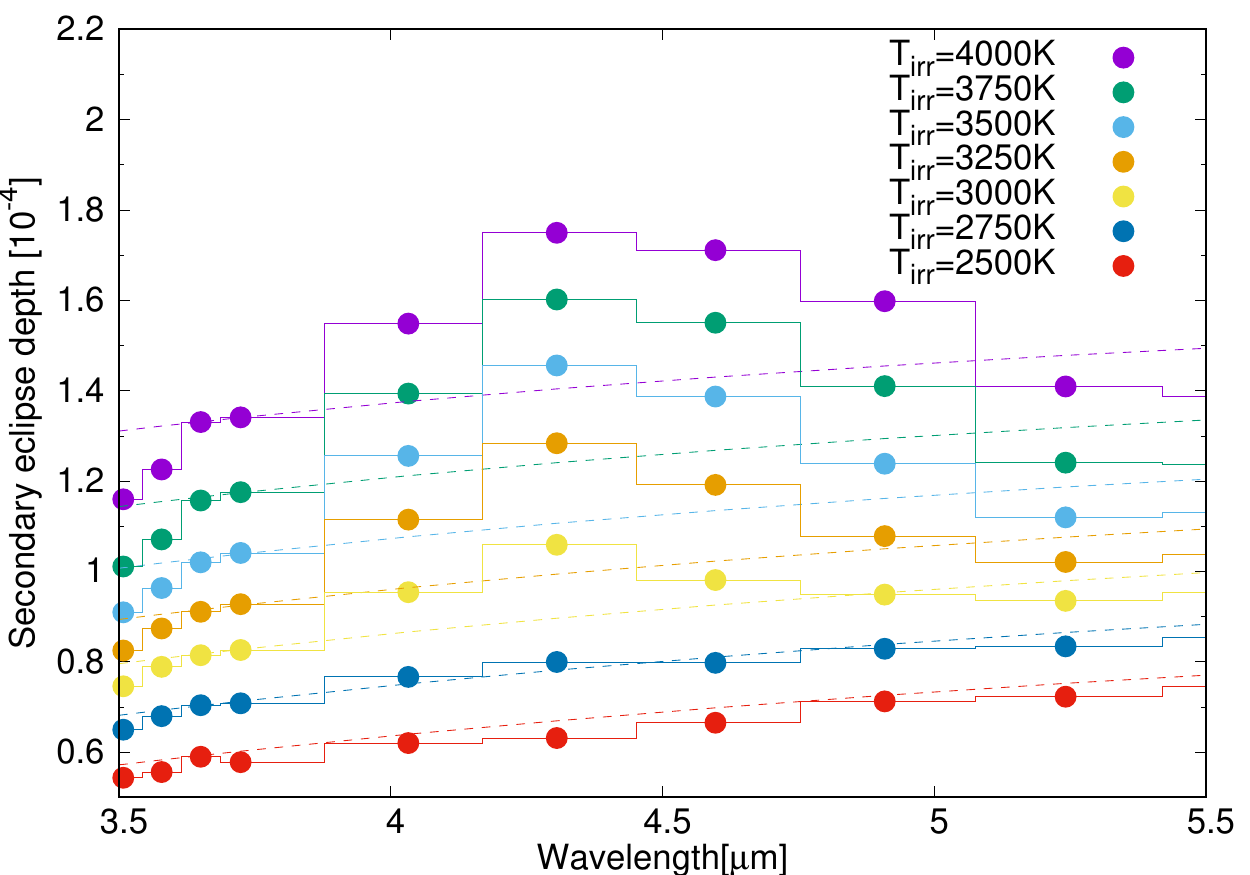}
    \end{center}
 \end{minipage}

\caption{Theoretically predicted spectra
of secondary eclipse depths of hot rocky exoplanets with radii of 2~R$_\oplus$ that have mineral atmospheres. The secondary eclipse depths with a high resolution (a) and Ariel resolution (b) are shown as a function of wavelength in the range of 3.5-5.5~$\mu$m. Seven substellar-point equilibrium temperatures are chosen: $T_{\mathrm{irr}}$ = 4000 K (violet), 3750 K (green), 3500 K (light blue), 3250 K (orange), 3000 K (yellow), 2750 K (blue) and 2500K (red) (see Eq. 2). In panel (b), the reference Black Body spectra, $BB_{ref}$, are shown by dotted lines (please see the definition of $BB_{ref}$ written at Sec.~\ref{sec:model}). 
}
\label{fig:tirr}
\end{figure}

Hot rocky exoplanets (HREs) with higher irradiation temperatures
could have more prominent SiO features, while it might be difficult to detect such features
for a 55 Cnc e like planet with Ariel, as demonstrated in Section~\ref{ssec:55}.
Here, we investigate the detectability of the SiO feature around 4~$\mu$m via secondary eclipse observations with Ariel for wide ranges of substellar-point equilibrium temperature $T_\mathrm{irr}$ and distance from the Earth. 
Figure~\ref{fig:tirr} shows the dependence of the 4-$\mu$m SiO feature 
on $T_{\mathrm{irr}}$. 
The SiO feature is found to become more prominent with $T_{\mathrm{irr}}$ in the spectra with a high resolution (Fig.~\ref{fig:tirr}a). 
In particular, with Ariel resolution (Fig.~\ref{fig:tirr}b), HREs of $T_{\mathrm{irr}}\geq3000$~K have 20-80 ppm level of the SiO features while the spectra for $T_{\mathrm{irr}}=2500$~K and 2750~K have features which are hardly discernible. This is because  the SiO partial pressure increases with $T_{\mathrm{irr}}$ and, as a consequence, changes the profile of the SiO emission feature, as also shown in previous studies \cite{Ito+2015}. Due to this, the optical thickness/brightness temperature of the 4-$\mu$m SiO feature is hardly different from the baseline surface blackbody, outside of the SiO feature, for $T_{\mathrm{irr}}=2500$~K. Also, given Ariel resolution, the feature is completely diminished for $T_{\mathrm{irr}}=2750$~K due to the small signal.

\begin{figure}
  \begin{minipage}{0.45\textwidth}
    ($a$) Observation time: 18.13 hours (5 eclipses) 
    \begin{center}
  \includegraphics[width=\textwidth]{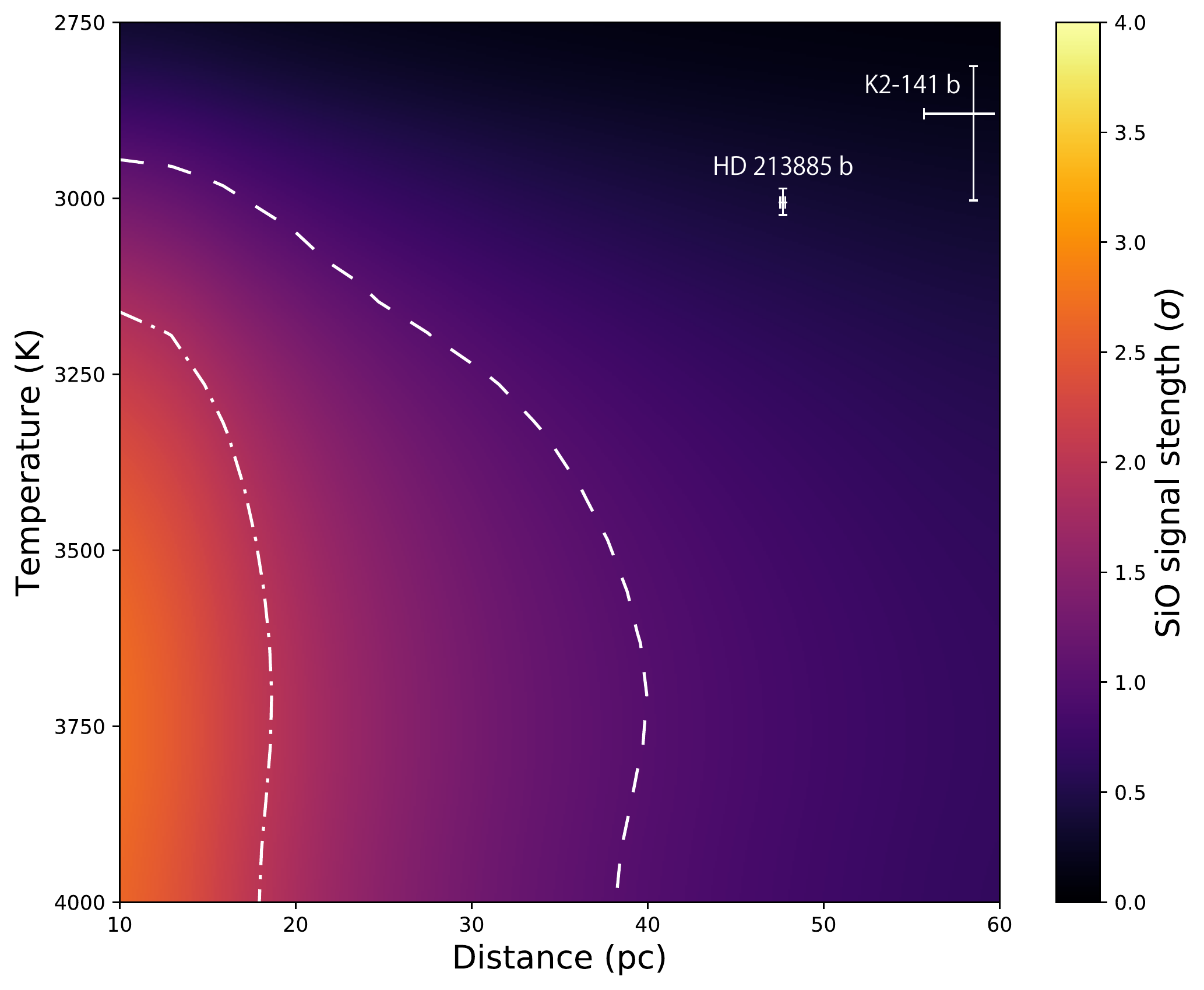}
    \end{center}
 \end{minipage}
     \begin{minipage}{0.45\textwidth}
    ($b$) Observation time: 36.25 hours (10 eclipses)
   \begin{center}
  \includegraphics[width=\textwidth]{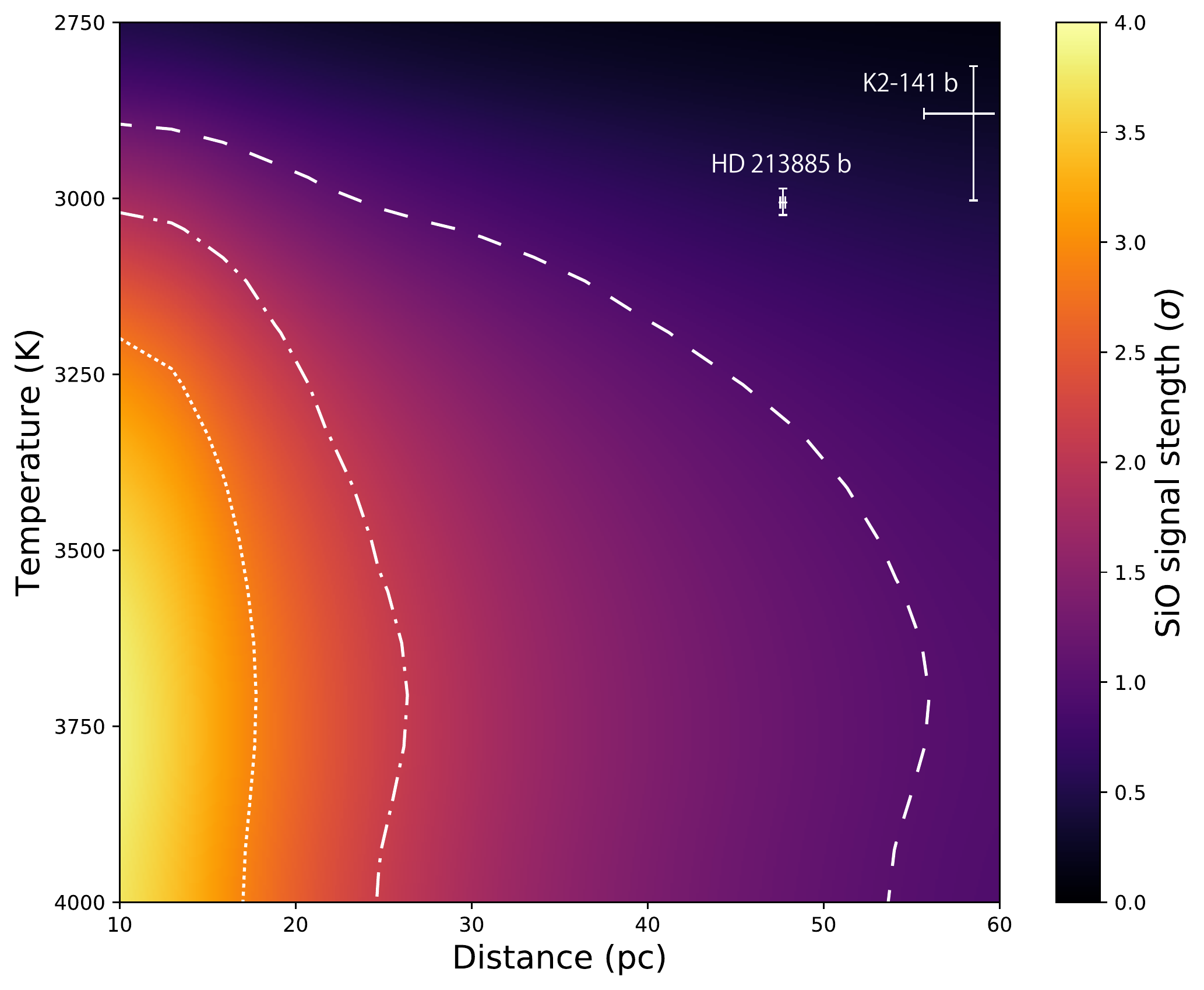}
   \end{center}
 \end{minipage}
\caption{Strength of the 4-$\mu$m SiO signal relative to the Ariel observational noise ($S_{\rm SiO}$; see Eq.~(\ref{eq:S_SiO})) within the secondary eclipse spectra of the mineral atmosphere of a hot rocky super-Earth of 2~R$_\oplus$ expressed as functions of substellar-point equilibrium temperature ($T_{\rm irr}$) and distance to Earth $d_\ast$ for  
18.13~h of observation (a) and 36.25~h of observation (b).
The $d_\ast$-$T_{\rm irr}$ relationships for $S_{\rm SiO}$ = 1, 2 and 3 are shown by 
dash, dotted dash and dotted curves, respectively.
The bars show the observed values of $d_\ast$ and $T_{\rm irr}$ of
K2-141~b \cite{Barragan+2018} and HD 213885~b \cite{Espinoza+2020}.
    }
\label{fig:grid}
\end{figure}

Figure~\ref{fig:grid} shows $S_{\rm SiO}$ for an HRE with radius of 2~R$_\oplus$ as a function of substellar-point equilibrium temperature ($T_\mathrm{irr}$) and distance to Earth ($d_\ast$) for the observation times stacked by 
18.13~hours (a) and 36.25~hours (b). 
Here we assume that the planet has an eclipse duration equal to that of 55 Cnc e (1.45 hours) and that Ariel observes 0.75 times this duration both before and after the eclipse (i.e., 3.63 hours is equivalent to one eclipse observation). 
Changing the eclipse duration would have little difference on the total time required to uncover the SiO signature: increasing the eclipse duration would decrease the error bar  per observation, thus requiring a fewer observations in total (and vice versa), keeping the total observation time roughly constant. For example, if one assumes circular orbits and edge-on to the line of sight (impact parameter of $b = 0$), 8, 9.2 and 10.3  eclipses of HREs with $T_{\mathrm{irr}}=3000$~K, 3500~K and 4000~K give about 14.5~hours as the sum of their eclipse duration 
 based on Eq.~(3) of \cite{Seager+2003}.

Since the SiO feature 
covers a broad range of wavelengths,
as shown in Fig.~\ref{fig:tirr}, 
we assume that if the signal deviates from 1 $\sigma$ (e.g., SiO top feature is outside the Ariel error bars), the SiO feature will be detectable from statistical analysis or retrieval techniques \cite{Al-refaie+2019}.
In Appendix,  we show that the SiO signal is actually detectable for $S_{\rm SiO} > 1$ using the retrieval techniques.
Thus, when $S_{\rm SiO} > 1$, we regard the SiO signal as detectable.

As also demonstrated in Fig.~\ref{fig:tirr}, 
the hotter the planets are, the more detectable the SiO signals are.
The stellar distance also plays an important role: the closer the star is, the better the detection is. 
When 18.13 hours of observations are stacked together (5 eclipses, see Fig.~\ref{fig:grid}a), it is possible to detect the SiO feature of HREs with 
$d_\ast$ of up to 40~pc and $T_\mathrm{irr}$ of at least 3000~K. Furthermore, when 36.25 hours of observations are combined, the SiO feature becomes more prominent,
as shown in Fig~\ref{fig:grid}b. 
In this case, it is possible to constrain SiO on HREs with $d_\ast$ of up to 55 pc. 
The detectability of SiO for HREs with radii 
different from 2~R$_\oplus$ are discussed in Section~\ref{ssec:dis1}.

Note that, there are no good targets for the SiO detection in the exoplanets detected so far. On the other hand, recent observations have reported the detection of new close-in rocky-density exoplanets such as K2-141~b \cite{Barragan+2018} and HD 213885~b \cite{Espinoza+2020} (see their observed values of $d_\ast$ and $T_{\rm irr}$ in Fig.~\ref{fig:grid}) which are hotter than 55~Cnc~e. Motivated by such an observational progress, the number of planets potentially having a detectable signature of mineral atmosphere are also discussed in Section~\ref{ssec:dis3}.

\section{Discussion} \label{sec:dis}
\subsection{Possibility to distinguish mineral atmosphere from volatile-rich atmosphere}\label{ssec:dis0}
\begin{figure}
\includegraphics[width=0.45\textwidth]{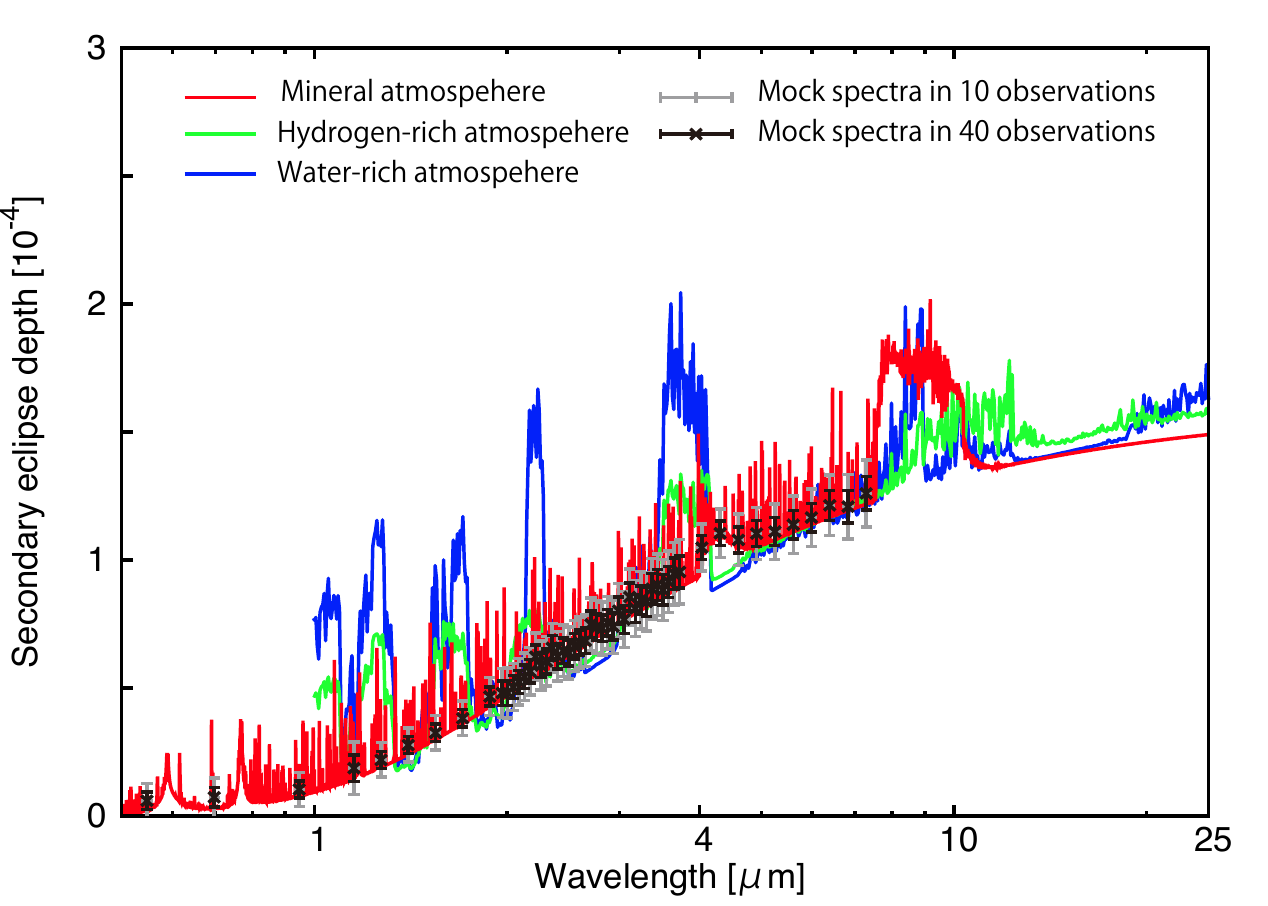}
\caption{
Theoretically predicted (red line) and mock (black and grey bars) spectra of secondary eclipse depth for a 55-Cnc-e-like transiting planet with a mineral atmosphere,
which are compared to theoretical spectra for hydrogen-rich (green line) and water-rich (blue line) atmospheres from Fig.~16 of \cite{Hu+2014}. 
The grey and black bars in the mock spectra represent cumulative errors estimated for 10 and 40 secondary eclipse observations with Ariel, respectively.
    }
\label{fig:55cnc_comp}
\end{figure}

The emission spectra of volatile-element-rich atmospheres are different from those of the mineral atmosphere \cite{Hu+2014,Lupu+2014}. 
Also, whether HREs have mineral atmospheres or atmospheres composed of highly volatile elements such as H and C is 
dependent upon 
their magma composition \cite{Schaefer+2009,Miguel+2011,Ito+2015,Schaefer+2012}.
Figure~\ref{fig:55cnc_comp} shows the difference between the secondary eclipse depth spectrum of the mineral atmosphere (red) and those of cloud-free hydrogen-rich (green) and water-rich atmospheres (blue) for a 55~Cnc~e like planet. The latter two spectra are predicted by the atmospheric model based on photo-/thermo-chemistry of hydrogen, carbon and oxygen \cite{Hu+2014}. 
It is found that there are some 
highly prominent features of 
H$_2$O, CO and CO$_2$ 
in the spectra of such volatile-rich atmospheres (see \cite{Hu+2014} for details),  
making 10 secondary eclipse observation with Ariel 
enough to distinguish the mineral atmosphere from the other two atmospheres. Also, while planets covered completely with thick clouds or with no atmosphere also show such flat spectra, detection of Na (0.6 $\mu$m) and K (0.8 $\mu$m) with ground-based telescopes would be helpful to distinguish the mineral atmosphere from such other possibilities.
Thus,  eclipse observations with Ariel would have a great potential to distinguish the mineral atmosphere from volatile-rich atmospheres. 
Such observations could give constraints on the bulk composition and formation processes of HREs. Future perspectives for characterisation of HREs is discussed in Section~\ref{ssec:dis3}.

\subsection{Scaling of SiO signal in HRE's size}\label{ssec:dis1}
Thanks to 
recent observational progresses, a growing number of Earth-size planets have been identified. 
In Section~\ref{ssec:ps} we investigated what kind of planets have potentially  detectable SiO features.
Although that estimation is for 2~R$_\oplus$ HREs, 
the signal-to-noise ratio can 
be approximately scaled by the square of planetary radius, 
because the mineral atmosphere is geometrically thin and the planetary gravity does not affect substantially the spectra.  
When calculating the spectra of HREs with substellar point equilibrium temperature of 3000~K for different radii of 2~R$_\oplus$ and 1~R$_\oplus$, except for the difference due to their radii, the spectral
difference between them turns out to be very small (at most 5~\%). 
Note that, the signal-to-noise ratio cannot be easily scaled for different stellar-types of host stars like that for planetary radii. This is because the spectrum of a host star affects the temperature profile and emission spectra of mineral atmosphere. 
If blackbody temperatures of 5000~K and 7000~K are assumed in our model for $T_\mathrm{irr}=3000$~K, in the planetary emission spectra, the 4-$\mu$m SiO signal strength is changed, only slightly (at most 10\%), but strong line features, such as the D-lines of Na and K, are changed significantly.

When the SiO signal-to-noise ratio shown in Fig~\ref{fig:grid} is scaled by the square of the planetary radius,
the SiO signal strength ($S_{\rm SiO}$) for HREs with radii of 1.4~R$_\oplus$  
is about 
half of that for 2~R$_\oplus$ HREs.
Thus, HREs with radii of 1.4~R$_\oplus$ are also potentially good targets to detect SiO in 36.25 hours of observation with Ariel, provided they are hotter than 3250~K and locate within 20~pc. 
Note that
scaling by the square of planetary radius would be invalid 
for Mars-size HREs because the gravity of such 
very small planets is so small 
that their atmospheric scale height is possibly comparable with 10 \% of their radii (i.e., geometrically thick). 
Then, the plane parallel approximation assumed in our model is also in valid. 
To simulate the emission spectra for such small HREs, a three-dimensional atmospheric model would be required, while the atmospheric observations for them would be more difficult than those for super-Earth-size HREs.

\subsection{How many HREs are potential targets?}\label{ssec:dis2}

While other currently known HREs are likely to be 
too faint for study with Ariel, ongoing and future exoplanet survey projects are expected to provide better targets for characterisation. The Transiting Exoplanet Survey Satellite (TESS) is predicted to find 280 exoplanets with radii of less than 2~R$_\oplus$ and analyses of data from the K2 mission continue  
to yield exoplanet detections \cite{Barclay+2018,Mayo+2018}. 
Here, we estimate the expected number of planets potentially having the detectable signature of mineral atmospheres with Ariel  observations.
Assuming a uniform distribution of G-type stars, the number of the detectable HREs, $N_{\mathrm{HRE}}$, is given by
\begin{equation}
    N_{\mathrm{HRE}} \approx r N_G \frac{4\pi}{3} d_{\mathrm{max}}^3
  \label{eq:num}
\end{equation}
where  
$r$ is the occurrence rate of HREs with $T_\mathrm{irr}\geq2900$~K hosted by G-type stars, $N_G$ is the number density of G-type stars and $d_{\mathrm{max}}$ is the limited distance to detect the signature of mineral atmospheres. 
The HREs with $T_\mathrm{irr}\geq2900$~K 
having the detectable SiO feature with reasonable observation time (Fig.~\ref{fig:grid}) corresponds to planets with orbital periods of $P\leq 1$~day. 
The recent occurrence rate estimate based on Kepler-data shows the occurrence rate of planets with $1.41 \leq R_p/R_{\oplus} \leq 2$ and $0.78  \mathrm{~days} \leq P \leq 1.56  \mathrm{~days}$ around G-type stars is 0.13 \% \cite{Kunimoto+2020}. 
Also, for such close-in super-Earths, the occurrence rate changes little over short orbital periods of $P  \leq 0.8  \mathrm{~days}$ \cite{Sanchis-Ojeda+2014}.
Here, we use $r=1.3\times 10^{-3}$ as a fiducial value of the occurrence rate.

In order to detect the 4-$\mu$m SiO feature, we estimated $d_{\mathrm{max}}$ to be $\sim$ 23~pc ($T_\mathrm{irr}=3000$~K), $\sim$ 44~pc ($T_\mathrm{irr}=3250$~K), $\sim$ 53~pc ($T_\mathrm{irr}=3500$~K), $\sim$ 57~pc ($T_\mathrm{irr}=3750$~K) and $\sim$ 53~pc ($T_\mathrm{irr}=4000$~K) with 36.25~hours of observation, as shown in Fig~\ref{fig:grid}. 
Also, the values of $d_{\mathrm{max}}$ for HREs with radii of 1.4~R$_\oplus$ are half of those of $2$R$_\oplus$ HREs, since the SiO signal strength for the former is half of that for the latter (see Section~\ref{ssec:dis1}). Thus, we take the simple mean $[(23^3+44^3+53^3+57^3+53^3)\times(1+0.5^3)/10)]^{1/3}\sim$ 40~pc as the fiducial value of the limited distance to detect SiO. When considering the observation to distinguish the mineral atmosphere from the volatile-rich atmospheres (Fig.~\ref{fig:grid}), one can simply double the distance to $80$~pc because the accuracy of 40~observation is twice larger than that of 10~observation (Fig.~\ref{fig:55cnc_distance}). Also, the number density of G-type stars, $N_G$, is known to be $\sim6\times10^{-3}$~pc$^{-3}$ (see a review by \cite{Traub+2010}).

Inputting $r=1.3\times 10^{-3}$, $N_G=6\times10^{-3}$~pc$^{-3}$ and $d_{\mathrm{max}}=40$~pc in Eq.~(\ref{eq:num}), we find that there are approximately two HREs whose 
SiO features are detectable with Ariel. 
Taking the transiting probability of $\sim$ 0.34 ($R_*/a$ for $T_\mathrm{irr}=3500$~K) into account, the expected number of HREs becomes about 0.6; that is, there may not be a potential target within 40~pc.
On the other hand, to distinguish the mineral atmosphere from the volatile-rich atmospheres via secondary eclipse observations with Ariel, 4 transiting HREs within 80~pc are expected as good potential targets. Thus, coming new HREs discovered by TESS and K2 will include good targets to characterize their atmospheres with Ariel observations.
Note that as the range of $T_\mathrm{irr}$ for the targets to identify volatile-rich atmospheres, which can have spectral features even if $T_\mathrm{irr}\leq2900$~K and their features are also easier to detect (see Fig.~\ref{fig:55cnc_comp}),  is wider than that of HREs with detectable SiO (for example, $T_\mathrm{irr}\sim2700$~K for 55 Cnc e), the estimated number is likely pessimistic.

\subsection{Near-future perspective for characterization of HREs}\label{ssec:dis3}
HREs would be the best targets for atmospheric observations to constrain their interiors with planned space missions because their secondary atmospheres are likely composed of materials directly vaporised from their magma ocean because of the rapid vaporisation/condensation (i.e, gas-melt equilibrium condition). 
Even if those planets are rocky, their interior structures and compositions are mostly unknown at present.

Some theoretical studies argue for the presence of not only terrestrial planets with interiors similar to those of solar system’s rocky planets, but also coreless planets \cite{Elkins-Tanton+2008} and carbon-rich exoplanets \cite{Madhusudhan+2012,Miozzi+2018}. However, it is difficult to determine uniquely the bulk composition only from the measured masses and radii, because exoplanet compositions with different materials can have similar densities \cite{Rogers+2010}. Additionally, it is interesting to note that some theoretical studies have predicted the mantle convection of rocky super-Earths like 55 Cnc e is very slow \cite{Tackley+2013,Miyagoshi+2018}. 
Such results suggest that the mantle and, therefore, magma of super-Earths possibly retain volatile materials because of delayed outgassing, 
while all of hydrogen and also water in  
their ancient atmosphere could have been evaporated away by strong UV irradiation from the host star \cite{Kurosaki+2014,Lopez2017}. 

The amount of highly volatile elements and the redox state of planetary magma are key factors for the vaporized atmospheric composition \cite{Schaefer+2009,Miguel+2011,Ito+2015,Schaefer+2012}. A coreless planet has a magma that is highly oxidized with abundant iron-bearing oxides, while a carbon-rich planet has a magma that is highly reduced with abundant SiC, pure carbon and Si. If the magma includes highly volatile elements, the atmosphere consists mainly of reducing gases such as CO, H$_2$S and CH$_4$ from the reduced magma, and of oxidizing gases such as CO$_2$, SO$_2$ and H$_2$O from the oxidized magma \cite{Schaefer+2012}. On the other hand, Na and SiO are the main atmospheric species on volatile-free magma \cite{Schaefer+2009,Miguel+2011,Ito+2015}. To remove such degeneracy in the interior composition of rocky exoplanets, atmospheric observations would be helpful giving additional constraints. 

Such a large variety of  
secondary atmospheres and 
interiors of HREs are theoretically expected, as mentioned above. Observational constraints of the atmospheres  of HREs with a dedicated space mission like Ariel would therefore be very  important. As we demonstrated in Section~\ref{ssec:dis2}, Ariel will be able to characterise the atmospheres of some HREs. It will open the new era of comparative planetology for rocky planets inside and beyond the solar system. Additionally, it is helpful to understand the planetary formation processes, especially planetary migration in proto-planetary disks \cite{Morbidelli+2016}. 
Also, 55 Cnc e can be characterized with Ariel observation while its host star is too bright for the observation with JWST. 

The thermal phase curve and transmission spectra of 55 Cnc e has been observed but its atmospheric composition is still debatable. The thermal phase curve observation found the large day-night temperature difference and 
significant eastward hot-spot shift \cite{Demory+2016}.  The presence of thick atmosphere is consistent with such features but thin atmosphere like a mineral atmosphere is not \cite{Zhang+2017,Hammond+2017}.  Motivated by this, some possibilities of 55 Cnc e's thick atmospheres such as Nitrogen-dominated atmospheres have been argued \cite{Miguel+2019,Zilinskas+2020}.
Also, 
the transmission spectra of 55 Cnc e have suggested the atmosphere contains a non negligible amount of light gases such as hydrogen \cite{{Tsiaras+2016}} but it does not include water vapour \cite{Esteves+2017}. The presence of abundant hydrogen but the absence of water vapour might suggest that the atmosphere is vaporized from reduced magma retaining hydrogen. If so, 55 Cnc e’s atmosphere might include not only H$_2$ but also CO and SiO \cite{Schaefer+2012}. The spectral features of such gas species are expected to be detectable with Ariel, as shown in Fig.~\ref{fig:55cnc_comp}.
On the other hand, a recent study has proposed a new scenario that 55 Cnc e is able to retain hydrogen if the planet became tidally locked before hydrogen-dominant atmosphere was photo-evaporated away \cite{Modirrousta-Galian+2020b}. According to the study, 55 Cnc e may host a significant amount of hydrogen at its terminator but the day-side may additionally have a mineral atmosphere. Even in such a case, although the detection of SiO would be difficult, Ariel observations could be helpful by ruling out other atmospheric compositions. In this case, the observation of a featureless emission spectrum would be conducive of a mineral atmosphere resulting from thermal emission of surface magma shown in Fig.~\ref{fig:55cnc_comp}.

\section{Conclusions} \label{sec:con}

In this study, we investigated under what conditions volatile-free hot rocky exoplanets 
potentially have detectable SiO features in secondary eclipse observations with Ariel. 
Our results demonstrate that the SiO emission feature of the mineral atmosphere around 4~$\mu$m is detectable for hot rocky super-Earths
with a irradiation temperature of 3000~K and a distance from Earth of up to 20~pc and for ones hotter than 3500~K and closer than 50~pc.
through 10 secondary eclipse observations. 
Also, in the case of 55~Cnc~e, we find that the detection of SiO would be difficult but 10 eclipse observation with Ariel would suffice to distinguish the mineral atmosphere from the cloud-free, hydrogen-rich or water-rich atmospheres.


\begin{acknowledgements}
We appreciate Dr. Renyu Hu for providing us with the calculation data of emission spectra of the hydrogen-rich and water-rich atmospheres that are used in Fig.~\ref{fig:55cnc_comp}.
This project has received funding from the European Union's Horizon 2020 research and innovation programme (grant agreement No. 776403, ExoplANETS A). 
Furthermore, we acknowledge funding by the Science and Technology Funding Council (STFC) grants: ST/K502406/1, ST/P000282/1, ST/P002153/1, and ST/S002634/1. 
\end{acknowledgements}

%
%



\begin{thebibliography}{}
%
%

\bibitem{Al-refaie+2019} Al-Refaie, A.~F., Changeat, Q., Waldmann, I.~P., Tinetti, G..\ TauREx III: A fast, dynamic and extendable framework for retrievals.\ arXiv e-prints arXiv:1912.07759. (2019)

\bibitem{Asimow+98} Asimow, P.~D., Ghiorso, M.~S.\ 1998.\ Algorithmic modifications extending MELTS to calculate subsolidus phase relations.\ American Mineralogist 83, 1127-1132.

\bibitem{Barclay+2018} Barclay, T., Pepper, J., Quintana, E.~V.\ A Revised Exoplanet Yield from the Transiting Exoplanet Survey Satellite (TESS).\ The Astrophysical Journal Supplement Series 239, 2. (2018)
\bibitem{Barragan+2018} Barrag{\'a}n, O., and 34 colleagues.\ K2-141 b. A 5-M$_{{\ensuremath{\oplus}}}$ super-Earth transiting a K7 V star every 6.7 h.\ Astronomy and Astrophysics 612, A95. (2018)

\bibitem{Bourrier+2018} Bourrier, V., and 11 colleagues. \ The 55 Cnc system reassessed.\ Astronomy and Astrophysics 619, A1. (2018)

\bibitem{Changeat+2019} Changeat, Q., and 3 colleagues. \ Towards a more complex description of chemical profiles in exoplanets retrievals: A 2-layer parameterisation.\ The Astronomical Journal 886 39. (2019)

\bibitem{Changeat+2020} Changeat, Q., and 3 colleagues. \ Impact of Planetary Mass Uncertainties on Exoplanet Atmospheric Retrievals.\ The Astrophysical Journal 896. (2020)

\bibitem{Changeat+2020b} Changeat, Q., and 6 colleagues. \ Alfnoor: A Retrieval Simulation of the Ariel Target List.\ The Astronomical Journal 160 80. (2020)

\bibitem{Demory+2016} Demory, B.-O. and 13 colleagues \ A map of the large day-night temperature gradient of a super-Earth exoplanet.\ Nature 532, 207-209. (2016)
\bibitem{Dragomir+2014} Dragomir, D., Matthews, J.~M., Winn, J.~N., Rowe, J.~F.\ New MOST Photometry of the 55 Cnc System.\ Formation, Detection, and Characterization of Extrasolar Habitable Planets 52. (2014)
\bibitem{Edwards+2019} Edwards, B., Mugnai, L., Tinetti, G., Pascale, E., Sarkar, S.\ An Updated Study of Potential Targets for Ariel.\ The Astronomical Journal 157, 242. (2019)
\bibitem{Ehrenreich+2012} Ehrenreich, D., and 11 colleagues .\ Hint of a transiting extended atmosphere on 55 Cnc b.\ Astronomy and Astrophysics 547, A18. (2012)
\bibitem{Elkins-Tanton+2008} Elkins-Tanton, L.~T., \& Seager, S., Astrophysical Journal, 688, 628 (2008)
\bibitem{Essack+2020} Essack, Z., Seager, S., Pajusalu, M. \ Low-albedo Surfaces of Lava Worlds.\ The Astrophysical Journal 898. (2020)

\bibitem{Espinoza+2020} Espinoza, N., and 62 colleagues.\ HD 213885b: a transiting 1-d-period super-Earth with an Earth-like composition around a bright (V = 7.9) star unveiled by TESS.\ Monthly Notices of the Royal Astronomical Society 491, 2982. (2020)
\bibitem{Esteves+2017} Esteves, L.~J., de Mooij, E.~J.~W., Jayawardhana, R., et al., Astronomical Journal, 153, 268 (2017)

\bibitem{Feroz+2006} Feroz, F., and 2 colleagues.\ MultiNest: an efficient and robust Bayesian inference tool for cosmology and particle physics. \ Nonthly Notices of the RAS 398: 1601-1614 (2009)

\bibitem{Fluton+2017} Fulton, B.~J., and 12 colleagues \ The California-Kepler Survey. III. A Gap in the Radius Distribution of Small Planets.\ The Astronomical Journal 154, 109. (2017)
\bibitem{Gardner+2006} Gardner, J.~P., and 22 colleagues.\ The James Webb Space Telescope.\ Space Science Reviews 123, 485. (2006)
\bibitem{Gillon+2012} Gillon, M., and 11 colleagues.\ Improved precision on the radius of the nearby super-Earth 55 Cnc e.\ Astronomy and Astrophysics 539, A28. (2012)

\bibitem{Ghiorso+95} Ghiorso, M.~S., Sack, R.~O. \ Chemical mass transfer in magmatic processes IV. A revised and internally consistent thermodynamic model for the interpolation and extrapolation of liquid-solid equilibria in magmatic systems at elevated temperatures and pressures.\ Contributions to Mineralogy and Petrology 119, 197-212. (1995)

\bibitem{Gordon+96} S. {Gordon} and B. J. {McBride} \ Computer Program for Calculation of Complex Chemical Equilibrium Compositions and Applications \ NASA Reference Publication 1311 (1996)

\bibitem{Hammond+2017} Hammond, M., Pierrehumbert, R.~T.\ 2017.\ Linking the Climate and Thermal Phase Curve of 55 Cancri e.\ The Astrophysical Journal 849. (2017)
\bibitem{Hu+2014} Hu, R., Seager, S.\ Photochemistry in Terrestrial Exoplanet Atmospheres. III. Photochemistry and Thermochemistry in Thick Atmospheres on Super Earths and Mini Neptunes.\ The Astrophysical Journal 784, 63. (2014)
\bibitem{Ito+2015} Ito, Y., Ikoma, M., et al.\ Theoretical Emission Spectra of Atmospheres of Hot Rocky Super-Earths.\ The Astrophysical Journal 801, 144. (2015)
\bibitem{Jin+2018} Jin, S., Mordasini, C.\ Compositional Imprints in Density-Distance-Time: A Rocky Composition for Close-in Low-mass Exoplanets from the Location of the Valley of Evaporation.\ The Astrophysical Journal 853, 163. (2018)
\bibitem{Kendrew+2015} Kendrew, S., and 15 colleagues.\ The Mid-Infrared Instrument for the James Webb Space Telescope, IV: The Low-Resolution Spectrometer.\ Publications of the Astronomical Society of the Pacific 127, 623. (2015)

	\bibitem{Kurosaki+2014} Kurosaki, K., Ikoma, M., \& Hori, Y., Astronomy \& Astrophysics, 562, A80 (2014)
\bibitem{Kunimoto+2020} Kunimoto, M., Matthews, J.~M.\ Searching the Entirety of Kepler Data. II. Occurrence Rate Estimates for FGK Stars.\ The Astronomical Journal 159, 248. (2020)

\bibitem{Lopez2017} Lopez, E.~D., Monthly Notices, 472, 245 (2017)

\bibitem{Lupu+2014} Lupu et al.\ The Atmospheres of Earthlike Planets after Giant Impact Events.\ The Astrophysical Journal 784, 27. (2014)
\bibitem{Lothringer+2018} Lothringer, J.~D., Barman, T., Koskinen, T.\ 2018.\ Extremely Irradiated Hot Jupiters: Non-oxide Inversions, H$^{-}$ Opacity, and Thermal Dissociation of Molecules.\ The Astrophysical Journal 866. (2018)

\bibitem{Madhusudhan+2012} Madhusudhan, N., Lee, K.~K.~M., \& Mousis, O.\ 2012, Astrophysical Journal Letter, 759, L40
\bibitem{Mayo+2018} Mayo, A.~W., and 27 colleagues \ 275 Candidates and 149 Validated Planets Orbiting Bright Stars in K2 Campaigns 0-10.\ The Astronomical Journal 155, 136. (2018)
\bibitem{Miguel+2011} Miguel, Y., Kaltenegger, L., Fegley, B., Schaefer, L.\ Compositions of Hot Super-earth Atmospheres: Exploring Kepler Candidates.\ The Astrophysical Journal 742, L19. 
\bibitem{Miguel+2019} Miguel, Y.\ \ Observability of molecular species in a nitrogen dominated atmosphere for 55 Cancri e.\ Monthly Notices of the Royal Astronomical Society 482, 2893-2901. (2019)

\bibitem{Miyagoshi+2018} Miyagoshi, T., Kameyama, M., \& Ogawa, M., Earth, Planets, and Space, 70, 200 (2018)
\bibitem{Morbidelli+2016} Morbidelli, A., Raymond, S.~N.\ Challenges in planet formation.\ Journal of Geophysical Research (Planets) 121, 1962. (2016)
\bibitem{Miozzi+2018} Miozzi, F., Morard, G., Antonangeli, D., et al., Journal of Geophysical Research (Planets), 123, 2295 (2018)
\bibitem{Mugnai+2019} Mugnai, L.~V., Pascale, E., Edwards, B., et al.\ 2020, ArielRad: the Ariel Radiometric Model .\ arXiv:2009.07824 
\bibitem{Modirrousta-Galian+2020a} Modirrousta-Galian, D., Ito, Y., Micela, G.\ 2020.\ Exploring Super-Earth Surfaces: Albedo of Near-Airless Magma Ocean Planets and Topography.\ arXiv e-prints.

\bibitem{Modirrousta-Galian+2020b} Modirrousta-Galian, D., Locci, D., Tinetti, G., Micela, G.\ Hot Super-Earths with Hydrogen Atmospheres: A Model Explaining Their Paradoxical Existence.\ The Astrophysical Journal 888. (2020)

\bibitem{Mugnai+2020} Mugnai, L., and 4 colleagues\ ArielRad: the Ariel Radiometric Model.\ Experimental Astronomy 50, 303-328. (2020)

\bibitem{Nelson+2014} Nelson, B.~E., and 7 colleagues .\ The 55 Cnc planetary system: fully self-consistent N-body constraints and a dynamical analysis.\ Monthly Notices of the Royal Astronomical Society 441, 442. (2014)
\bibitem{Owen+2017} Owen, J.~E., Wu, Y.\ The Evaporation Valley in the Kepler Planets.\ The Astrophysical Journal 847, 29. (2017)
\bibitem{Rogers+2010} Rogers, L.~A., \& Seager, S., Astrophysical Journal, 712, 974 (2010)
\bibitem{Sanchis-Ojeda+2014} Sanchis-Ojeda, R., Rappaport, S., Winn, J.~N., Kotson, M.~C., Levine, A., El Mellah, I.\ A Study of the Shortest-period Planets Found with Kepler.\ The Astrophysical Journal 787, 47. (2014)
\bibitem{Schaefer+2009} {Schaefer}, Schaefer, L., Fegley, B.\ Chemistry of Silicate Atmospheres of Evaporating Super-Earths.\ The Astrophysical Journal 703, L113. (2009)
\bibitem{Schaefer+2012} Schaefer, L., Lodders, K., Fegley, B.\ Vaporization of the Earth: Application to Exoplanet Atmospheres.\ The Astrophysical Journal 755, 41. (2012)
\bibitem{Seager+2003} Seager, S., Mall{\'e}n-Ornelas, G.\ A Unique Solution of Planet and Star Parameters from an Extrasolar Planet Transit Light Curve.\ The Astrophysical Journal 585, 1038. (2003)
\bibitem{Tackley+2013} Tackley, P.~J., Ammann, M., Brodholt, J.~P., et al., Icarus, 225, 50 (2013)
\bibitem{Tinetti+2018} Tinetti, G., and 243 colleagues.\ A chemical survey of exoplanets with Ariel.\ Experimental Astronomy 46, 135. (2018)
\bibitem{Toon+89} Toon, O.~B., McKay, C.~P., Ackerman, T.~P., Santhanam, K.\ Rapid calculation of radiative heating rates and photodissociation rates in inhomogeneous multiple scattering atmospheres.\ Journal of Geophysical Research 94, 16287-16301. (1989)
\bibitem{Tsiaras+2016} Tsiaras, A., Rocchetto, M., Waldmann, I.~P., et al., Astrophysical Journal,, 820, 99 (2016)
\bibitem{Traub+2010} Traub, W.~A., Oppenheimer, B.~R.\ Direct Imaging of Exoplanets.\ Exoplanets 111. (2010)
\bibitem{Valencia+2010} Valencia, D., Ikoma, M., Guillot, T., Nettlemann, N.\ Composition and Fate of Short-Period Super-Earths: The Case of CoRoT-7b.\ Lunar and Planetary Science Conference 1872. (2010)
\bibitem{Zhang+2017} Zhang, X., Showman, A.~P.\ Effects of Bulk Composition on the Atmospheric Dynamics on Close-in Exoplanets.\ The Astrophysical Journal 836. (2017)
\bibitem{Zilinskas+2020} Zilinskas, M., Miguel, Y., Molli{\`e}re, P., Tsai, S.-M. \ Atmospheric compositions and observability of nitrogen-dominated ultra-short-period super-Earths.\ Monthly Notices of the Royal Astronomical Society 494, 1490-1506. (2020)


\end{thebibliography}


\begin{figure*}
\includegraphics[width=0.98\textwidth]{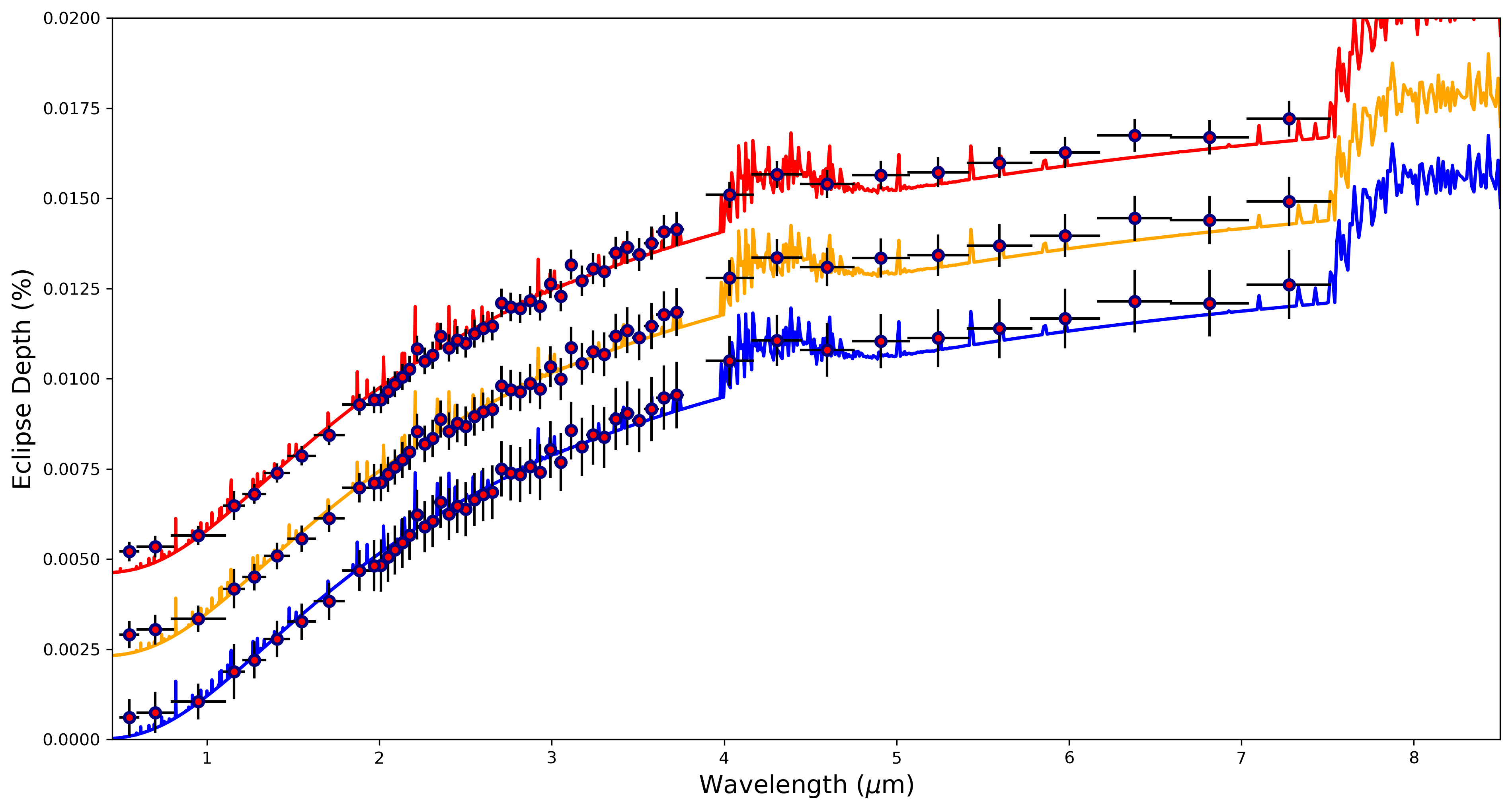}
\includegraphics[width=0.38\textwidth]{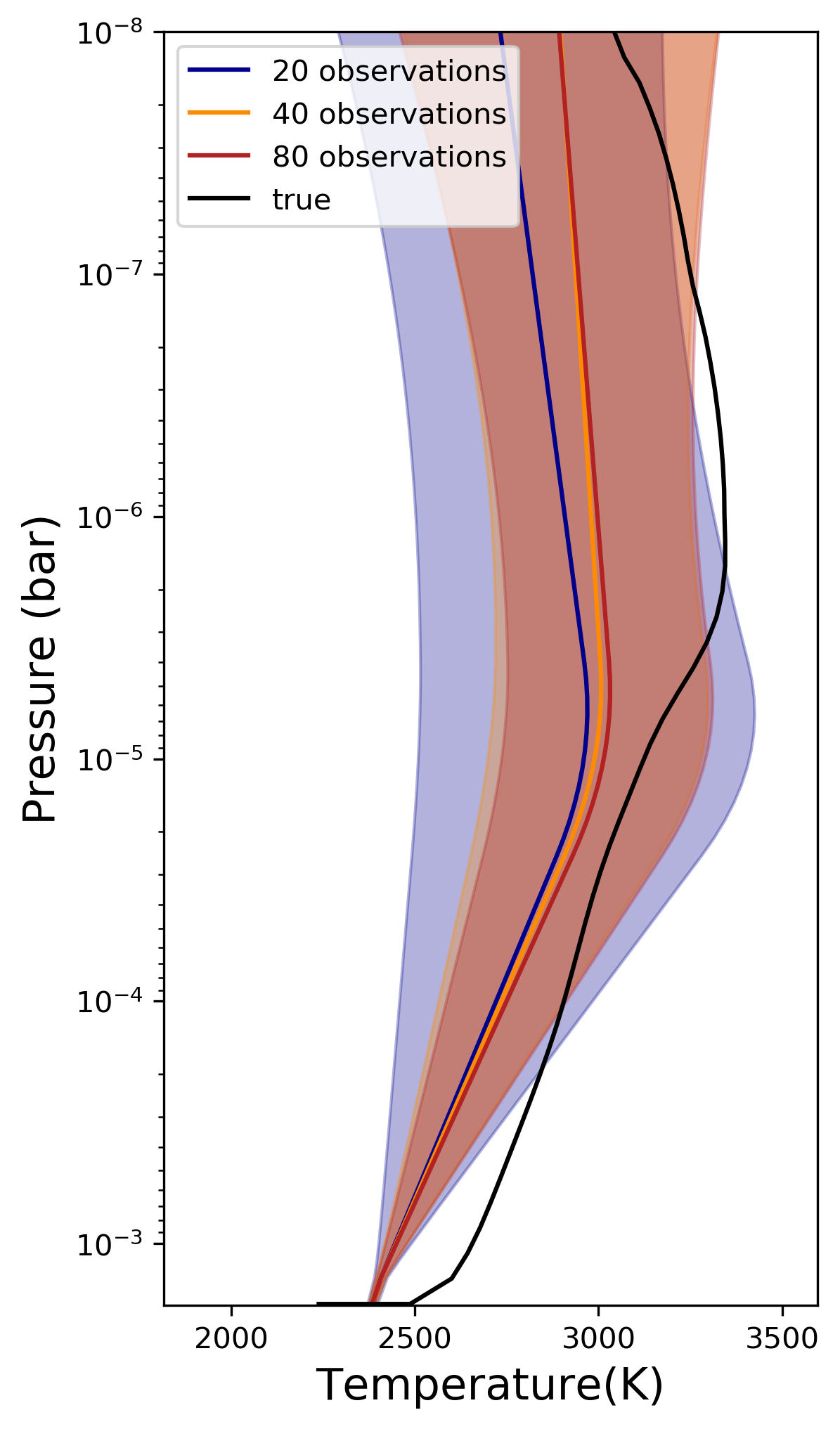}
\includegraphics[width=0.62\textwidth]{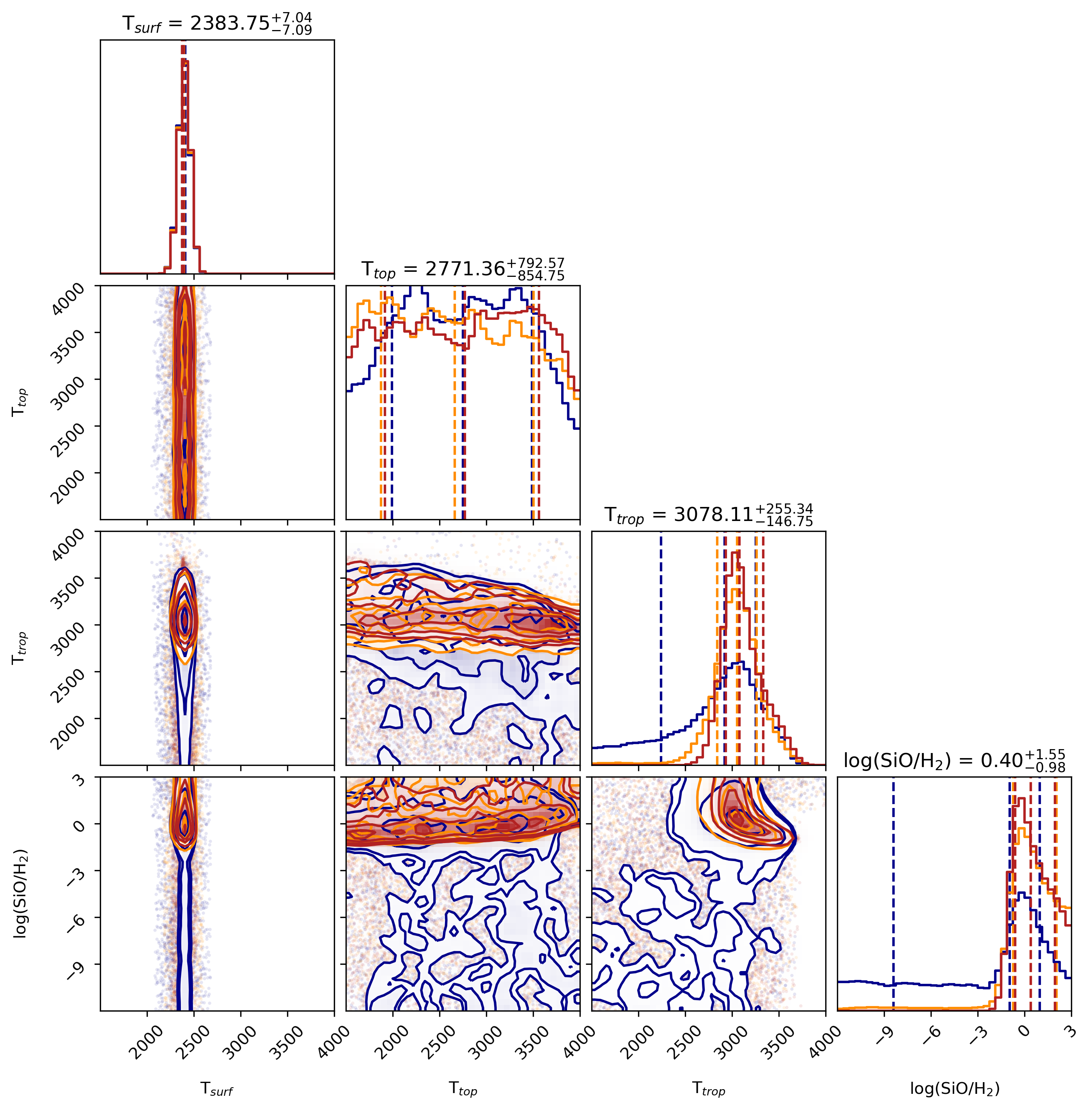}
\caption{Results of our retrieval analysis on the eclipse spectra of a mineral atmosphere on 55-Cancri\,e assuming 20 (blue), 40 (orange) and 80 (red) visits with Ariel. Top: Simulated Ariel observations and best fit spectra; Bottom Left: Retrieved mean and 1$\sigma$ temperature profiles; Bottom Right: Posterior distributions of the free parameters.  
}
\label{fig:retrieval}
\end{figure*}

\section*{Appendix A: 55-Cancri\,e retrieval simulations}

In order to investigate the feasibility to detect SiO with future telescopes, we simulate the atmosphere of a 55-Cancri\,e like planet using the model from \cite{Ito+2015} and estimating the Ariel noise with the Ariel Radiometric Model (ArielRad) from \cite{Mugnai+2020}. The observations are assumed at Tier-2 resolution \cite{Edwards+2019}\cite{Changeat+2020b} and we investigate the combined transits of 20, 40 and 80 visits. The corresponding S$_{SiO}$ signal strength for these visits can be found in Figure \ref{fig:55cnc_distance} and we note that a value of 1$\sigma$ is obtained between 20 and 40 combined transits. We then performed an atmospheric retrieval using TauREx3 on the 3 cases, assuming a plane parallel atmosphere with 100 layers up to 10$^{-5}$ Pa. The surface pressure was fixed to its true value. Since a surface pressure and the molar fraction of a gas have the same effect on its optical depth, only the partial pressure of the gas inducing a spectral signature can be retrieved from atmospheric spectra in principle. Note that, however, the surface pressures of mineral atmospheres can be determined from our atmospheric model using retrieved surface temperature.

Since the Ariel spectra obtained for a mineral atmosphere have a relatively low information content, we fit the simulated spectra using a simplified retrieval model. The planetary radius and mass were fixed \cite{Changeat+2020} to the literature values as more accurate constraints can be obtain from Radial Velocity and Transit measurements. For the temperature structure, we retrieved a heuristic profile comprised of 3 freely moving temperature-pressure points located at the surface, at 1 Pa and 10$^{-5}$ Pa. The atmosphere was assumed to be composed of H$_2$, He and SiO with the molecular ratio $X_{He}/X_{H_2}$ fixed to solar values and the ratio $X_{SiO}/X_{H_2}$ being the only free parameter of the chemistry.
To explore the parameter space we use the Nested Sampling algorithm MultiNest \cite{Feroz+2006} with 1000 live points and an evidence tolerance of 0.5.

From those retrievals, we find that the SiO spectra feature at 4.5$\mu$m is difficult to capture with 20 combined Ariel observations. For this case, the posterior distribution shows hints of the SiO signal, but a large tail is observed towards the low abundances, which would not allow to definitively conclude for this case. In the 40 and 80 observations cases, however, the noise is greatly reduced and a clear lower limit on the molecular ratio is observed (log SiO/H$_2$ = 0.4$^{+1.5}_{-1.0}$). While the precise abundances can't be obtained, a retrieval analysis would give strong indications in favor of a mineral atmosphere. The retrieved temperature structure for this example follows the input profile, but large differences are noticeable due to the differences between the forward and retrieval models. This is known to lead to biases that could potentially be mitigated when interpreting the results using self-consistent models or replacing the retrieval model with a more realistic scenario \cite{Changeat+2019,Changeat+2020b}.  We note that for hotter planets, the detection of an SiO signal with Ariel would be much easier. This is because the 4-$\mu$m-SiO signals deviate from $BB_\mathrm{ref}$ at three or four Ariel wavelength-bins for hotter planets (Fig.~\ref{fig:tirr}), while it deviates from $BB_\mathrm{ref}$ at the two bins ($\sim4.0\mu$m and $\sim4.3\mu$m) for the 55 Cnc e case. On top of this, when comparing with other models for the magma and atmosphere composition (see Figure \ref{fig:55cnc_comp}), the mineral atmosphere case appears as the worst case scenario since bigger features are observed in the cases of Hydrogen rich or Water rich atmosphere. In practice, it is likely that much less than 20 visits would be need for 55-Cancri\,e like planet to rule out the Hydrogen and Water rich cases.

\end{document}